\documentclass[12pt]{article}
\usepackage{epsfig}

\title{Conceptual problems in phenomenological interpretation in searches
for variation of constants \& violation of various invariances }

\author{Savely G. Karshenboim\\
\small \em D. I. Mendeleev Institute for Metrology (VNIIM),
St.~Petersburg 190005,
Russia\\
\small \em and Max-Planck-Institut f\"ur Quantenoptik, 85748 Garching, Germany\\
\small Email: sek@mpq.mpg.de}
\begin{document}

\maketitle

\begin{abstract}
At present a number of current or proposed experiments are
directed towards a search for a `new physics' by detecting variations of
fundamental physical constants or violations of certain basic
symmetries. Various problems related to the phenomenology of such
experiments are considered here.
\end{abstract}

\section{Introduction}

Progress in precision studies and shortage of data on possible
extension of the Standard Model of weak, electromagnetic and
strong interactions have produced a situation when a number of
experiments to search for so-called `new physics' have been
performed or planned
in atomic physics. Among such experiments are a search for an
electric dipole moment of an electron and a neutron, search for
variation of fundamental constants and violation of Lorentz
invariance etc.

While a number of experiments are designed to check a particular
theory, the others have aimed to look for `new physics' in a
`model-independent way'. Most of such experiments involve various
constraints even within a phenomenological interpretation. Such
conceptual problems of new physics in phenomenological
interpretation are considered in this note.

An example of such a problem is a relation of possible time or
space variation of fundamental constants and a basic relativistic
principle of local position/time invariance
(LPI/LTI)\footnote{Theoretically, LTI is indeed a part of LPI, but
from the experimental point of view the related experiments are
completely different.}. Some scientists consider possible
variation of constants as violation of LTI. However, that is not
correct.

We should acknowledge that there are two basic possibilities of
variations of constants. One is a result of certain violation of
LPI, while the other is an observational effect of the interaction
with environment, such as the bath of Cosmic Microwave Background
(CMB) radiation of photons, neutrinos, gravitons, Dark Matter
(DM), matter etc.

The idea of scaling of the environment is changing once we
increase accuracy. For example, we can say that the Earth gravity
at accuracy better than one-ppm level is described by three
forces: attraction by Earth, Sun and Moon and the acceleration of
free fall, $g$, is a parameter of the interaction with Earth only.
Alternatively, we can say that the complete gravitational force is
always $mg$ and it is varying in time because of the relative motion
of Earth, Sun and Moon. That is not only a matter of definition.
It depends on natural time scale of the experiment with respect to
the periods of Earth motion and on whether we understand
the planetary motion properly.

Interactions with the environment can always have a slow
component, which is in a way universal. From the observational point
of view a slowly changing parameter presents a kind of variation
of a constant, but indeed that is not violation of LTI.

As a matter of fact, we have to acknowledge that the most popular
model of evolution of the universe suggests so-called inflation
\cite{Linde} which is caused by a phase transition. The latter is
a transition between two phases of vacuum. In the former phase the
electron mass was zero and the proton mass was approximately
5-10\% lower than now.

Another example of conceptual problems is the electric dipole moment
(EDM) of an electron. The EDM of an electron can be caused by two
effects. One is some violation of the CP invariance in one or
other way, which allows a correlation between directions of a
vector (the EDM) and a pseudovector (the spin, ${\bf s}$). All
experiments have been interpreted in such a way. Meanwhile, there
is another opportunity due to a possible violation of the Lorentz
invariance. Such a violation can deliver a preferred frame (e.g.,
related either to the isotropy in CMB or to the local DM motion)
and a preferred direction related to our velocity ${\bf v}$ with
respect to the preferred frame. The violation could induce a
certain EDM directed with this preferred direction (along ${\bf
v}$) or in a direction of $\bigl[{\bf v}\times{\bf s}\bigr]$. All
the experiments have been treated in the former way, while
their results were considered as a model-independent limitation on
the electron's EDM.

\section{Variation of constants}

Here we consider various aspects of a possible interpretation of
experiments on variation of constants.

\subsection{Can constants vary?}

First of all, we note that the very motto `variation of constants'
is a jargon which is not related to reality. In a sense, the
constants cannot vary, because the idea of variation of the
constants is based on an assumption, that we can apply
conventional equations, but claim that some their parameters are
now adiabatically time- or space- dependent. That cannot be
correct. We should completely change the equations.

Let us show a simple example why we should. Consider a quite
unrealistic problem: a free electron in rest and the Planck
constant changing in time ($\hbar=\hbar(t)$), but not in space in
a particular frame. That is an isotropic situation and the angular
moment $L_z$ is conserved. On the other hand, it is equal to $\pm
(1/2) \hbar(t)$. Something should be changed. Applying
conventional equations of quantum mechanics to physics with
changing $\hbar$ produces an obvious contradiction. Another
example is considered in Sect.~\ref{moon}.

\subsection{Alternating the basic description}

At present, we have many equivalent ways to describe quantum
mechanics. Introduction of the `varying constants' into these
different descriptions produces different effects.

For example, even in classical mechanics we have equations (like
Newton's) and the variational principle of the least action. If we
introduced something similar to varying constants into the
Lagrangian we will change the equations. Similar dilemma exists in
quantum theory, where we should decide what the fundamental
description is and what its consequences are.

We believe \cite{looking} that the fundamental original
description is the Feynman's continual integral in quantum field
theory \cite{feynman} (which is similar to the path integral in
quantum mechanics).

Indeed, this integral is not well-defined, but it reproduces the
very soul of the quantum physics---the interference as its
backbone. When we do experiments, like the famous Young's
double-slit experiment, we look for an interference pattern. When
Bohr first created a theory of the hydrogen atom, he was also
dealing with interference effects such as standing waves. The
Schr\"odinger equation has taught us that relation for a photon
$E=\hbar\omega$ has deep meaning and in a sense any energy (at
least for a stationary problem) is certain frequency because the
stationarity is supported by a periodic motion and a resulting
standing wave. Having in mind the path integral, we see that a
stationary motion survives because of constructive interference
after going the same loop again and again, while for any other
frequency the interference is destructive. The Borh's orbit is a
kind of a standing wave. The least action principle (the least
distance Ferma principle for light) is also understood via a
constructive interference---the phase around the least phase
trajectory is almost not changing and that greatly enhances the
classical trajectory by the summation of the huge number of very
close trajectories, while far away from it the phase has no
minimum and changes linearly with the variation of any parameter
of the trajectory. The fast oscillating sum goes down because of
destructive interference of nearby trajectories.

To our mind the most natural way to go beyond conventional physics
phenomenologically is to change the phases in the continual
integral (which is the action) and to derive proper equations from
it.

Note, that to introduce medium into the Maxwell equation one has to
change the electromagnetic Lagrangian in a relatively easy way
(and that is related to a simple change in the action), while the
derived equations already involve various complications such as
derivatives.

\subsection{Gradient terms}

We need to emphasize that we deal with a phenomenological
description and do not discuss how various time- and
space-dependent factors can appear in the Lagrangian. We just
like to describe a proper framework for the interpretation of the data.

Introducing time- or space-dependent factors into the Lagrangian we
should expect their appearance in equations together with
their gradients. In principle, some gradients could appear even in
the Lagrangian. Once we have a derivative such as $\partial
Y/\partial x_0$, where $x_0=ct$ and $c=c(x)$, we have to think
which of many ways should be taken to modify the derivative:
\begin{eqnarray}
\frac{\partial Y}{\partial x_0}&\to&\frac{1}{c}\frac{\partial Y}{\partial t}\;,\nonumber\\
\frac{\partial Y}{\partial x_0}&\to&\frac{\partial Y}{\partial
(ct)}\;,\nonumber\\
\frac{\partial Y}{\partial x_0}&\to&\frac{\partial 1}{\partial
t}\frac{Y}{c}\;,\nonumber
\end{eqnarray}
with some of them including the derivative $\partial c/\partial t$.
But even if the Lagrangian in proper variables is free of
derivatives, the equations of motion derived from the Lagrangian
should contain some gradients.

Indeed, we should not be surprised by appearance of the time-
and/or space-gradients. We have a well-known self-consistent
example of a violation of the relativistic invariance with the
speed of light depending in principle on frame, location, time
etc. That is electrodynamics in media with a proper medium
density. When we consider it in a conventional way, four
three-dimensional field vectors have to be applied: ${\bf E}$,
${\bf D}$, ${\bf B}$ and ${\bf H}$. If we consider that as a
theory of photons, we should introduce, instead of the four former
field vectors, one four-dimensional vector $A_\mu$. When
presenting all the equations in such a way, we immediately arrive
at equations which contain not only two functions $\epsilon(x)$
and $\mu(x)$ (or even two tensor functions in a general case) but
also their derivatives. If afterwards we will try to describe the
complete interaction of two moving charges, their interaction
should also include the derivatives of $\epsilon(x)$ and $\mu(x)$.
That is far different from the naive expectation that it is enough
to write $c(x)$ instead of $c$ for the Lorentz force.

\subsection{Dimensional quantities}

Presence of the gradient terms allows to search for possible
variations related to dimensional quantities. However, even
without any gradient terms present in the basic equations, a
variation of dimensional quantities can be detected.

Any measurement is a comparison and we have to deal with a
`measured value' and a `reference value' of the same dimension.
However, applying some differential methods, a key comparison can
be performed between two values of the same quantity, e.g., the speed
of light, related to different directions. In other words, we can
look for a relative change of a certain dimensional quantity.

Note that such a statement is related not only to fundamental
physics, but also to practical issues. For example, the definition
of the kilogram states that the mass of the prototype is exactly one
kilogram. However, that does not mean that the mass of the
prototype is not changing. A change in the mass of a particular
object can be determined without any definition of the units. That
can be done in relative units and only needs definitions of the
involved quantities. Meanwhile, the definition of the kilogram
fixes only the numerical value of the mass of the prototype, while
the unit can change and that is detectable.

\subsection{Three kinds of searches}

Returning to possible variation of the constants, we conclude that
there are three basic kinds of searches \cite{looking}, which are
\begin{itemize}
\item a series of `fast' measurements with `long'
separation periods;
\item a `long' monitoring experiment;
\item a selectively sensitive `gradient experiment'.
\end{itemize}

Selective `gradient experiments' are model-dependent. They assume
a certain possible effect and look for it. That may be a search
for a gradient term, or a differential experiment. Indeed, to
isolate a particular effect one should have a model and thus all
such gradient terms are model-dependent and the dependence
sometimes goes much further than expected naively. Due to the
Lorentz invariance somewhat below the famous
Michelson-Morley experiment is discussed as an experiment of this kind.

Most of the experiments are of a different kind: they look for
certain values and check whether they change or not. However, a
crucial point is duration of different phases of the experiment.
In principle, there are two phases: reading the values (the
measurement proper) and the accumulation of the effect of the
variation (during a separation between the measurements).

The problem is that the accumulation period for a change of
parameters of a system, like, e.g., the electron's mass and
charge, usually does not involve any effects of the gradient
terms. The latter are important only during the measurements. If
the measurement\footnote{The measurements in sense of quantum
mechanics is determined by the interaction time and by the
coherence time. The measurement in sense of reading data consists
of a session of many `quantum measurements'. We indeed mean here
the duration of the quantum measurements.} is fast enough, the
effects of the gradients can be neglected and the most important
effect is the evolution of the parameters of the system under
study between the measurements, namely during the accumulation
time. In this case we can consider the same equations but with
varying parameters and do a model-independent evaluation. A
typical example of such kind of experiments is a study of the
variability of the constants by means of atomic physics. Since the
gradient are not involved the constraint can be achieved in such
experiments only on variations of dimensionless quantities.

Another situation occurs for various space tests of general relativity
and related experiments for $\partial G/\partial t$. The
accumulation time and the reading-data time is essentially the
same. Even if we try, similarly to atomic physics, to perform
brief, say, one-day measurements every year, even that would not
help. The problem is that when looking for a `rotation' of an
electron we cannot measure the phase of the rotation and deal with
something related in sense of classical physics to average
parameters of the orbital motion. The planetary motion allows us
to look for the phase of the rotation and thus the `coherence
time' is equal to many periods of evolution. In a sense that is
similar to the Ramsey method with two coherent space-separated
short measurements. As a result, the effects of the gradients are of
the same importance as effects of the time dependence of
conventional terms. Any interpretation of the data in such a case
is indeed strongly model-dependent.

Dealing with average values for planetary and atomic motion is
also not the same. The atomic orbits are quantized. Their
parameters do not depend on initial conditions, which determine
only probability to create one or other atomic state. Planetary
motion depends strongly on the initial conditions and thus even
experiments on average values have a kind of history and
accumulate effects from gradients.

Involvement of the gradient terms allows of constraining
variations of dimensional constants.

\subsection{An example\label{moon}}

Let us illustrate the consideration above with a clear example. We
consider a case of a non-relativistic classical problem of a
two-body system with one mass, $M$, much heavier than the other $m$
(e.g., a Sun--planet, or planet--satellite system). We neglect all
corrections in the order $m/M$ and for further simplicity suggest a
circular orbit.

The main parameters of the problem are: heavy mass $M$, lighter
mass $m$, gravitation constant $G$, orbital radius $R$ and orbital
velocity $v$. Starting with the equation of motion in a conventional
case ($G$, $M$ and $m$ do not depend on time) we find:
\begin{equation}\label{eq_mot1}
m \,{\bf a} = - \frac{GmM}{R^2}\, \frac{\bf R}{R}\;,
\end{equation}
or
\begin{equation}\label{eq_mot2}
{\bf a} = - \frac{GM}{R^2}\, \frac{\bf R}{R}\;,
\end{equation}
with appropriate initial conditions for a circular orbit.

If we assume that all the constants ($G$, $M$ and $m$) depend on
time, but expect that we can apply an adiabatic approximation,
i.e., neglect all time derivatives, we can still use the
equation of motion (\ref{eq_mot1}) or (\ref{eq_mot2}) and the
lighter mass, $m$, vanishes there. That means in particular
that any time dependence of $m$ is unimportant, because for slow
changes it looks natural to neglect all time gradients. In
particular, measuring the distance $R$ as a function of time we find
\begin{equation}\label{eq_R1}
R = \frac{GM}{v^2}\;.
\end{equation}
We note that the acceleration is orthogonal to the velocity and
thus the velocity has only a tangential component ($v=v_{||}$) which
is conserved. In other words, for time-depending terms we find a
proportionality law
\begin{equation}\label{eq_R1a}
R(t) \sim G(t)\,M(t)\;.
\end{equation}

However, the `adiabatic approximation' is inconsistent. Let us
consider the problem adiabatically, but neglecting the
time-gradients in conservation laws, not in the equation of
motion. As a result we find that (\ref{eq_R1}) is still correct;
however, the tangential velocity is not conserved. Instead, the
tangential component of the momentum
\begin{equation}\label{p_para}
p_{||} = m  v_{||}
\end{equation}
is. As a result we find
\begin{equation}\label{eq_R2}
R = \frac{GMm^2}{p_{||}^2}
\end{equation}
and for time-depending quantities the proportionality law takes
the form
\begin{equation}\label{eq_R2a}
R(t) \sim G(t)\,M(t)\,\bigl(m(t)\bigr)^2
\end{equation}
to be compared with (\ref{eq_R1a}).

The difference between (\ref{eq_R1a}) and (\ref{eq_R2a}) is caused
by a gradient term ${\bf v} \partial m/\partial t$ to appear in
the equation of motion (\ref{eq_mot1}). In other words, adiabatic
treatment of the conservation laws suggests a non-adiabatic
approximation in the equation of motion. The example shows that
the gradient terms in the equation of motions may be as important
as adiabatic effects---the former lead to $m^2$ in (\ref{eq_R2a}),
while the latter are responsible there for $M$.

The equations achieved above via the conservation laws do not
contains gradients directly. Still they allow to constraint
dimensional quantities. Technically that appears as a consequence
of presence of certain dimensional conserved quantities, such as
$p_{||}$ (as we demonstrated its conservation is related to a
gradient term of the equation of motion). The dimensionless
combination on which the constraint is achieved contains such a
conserved quantity. Technically it originates from the initial
conditions and that is why there is no analog of it in quantum
theory where the atomic energy levels are determined only by the
fundamental constants and not by any initial conditions.

One more point about this example is that the equation
(\ref{eq_R2a}) cannot be the end of the story. In the framework
given the mass should be conserved. There are two natural options
to describe the time dependence of the mass and both imply further
modifications of the scaling laws for $R(t)$.

The first idea is to allow a time dependence in the framework of
classical non-relativistic physics. An obvious mechanism is to
suggest that there is a mass in the space (e.g., dust particles)
which is not observable and the very presence of this mass allows
a change in the object moving through. This model is very similar
to introducing, e.g.,  the internal (thermodynamic) energy into the
mechanical consideration at the moment when it was absolutely unclear
what the substance is. A similar successful idea was to suggest
a neutrino to solve the problem of shortage of energy in the beta
decay.

Suggesting such a mechanism solves the problem of a possible time
dependence offering a mass transfer between unobservable dust and
a moving body. Meanwhile, it opens a question of a possible
transfer of momentum, angular momentum and kinetic energy. Any
particular model of the mass transfer sets constraints on the
transfers of other quantities and will produce different
corrections to (\ref{eq_R2a}). Note also that such a description will
require the introduction of some functions to describe the dust
particle in continuous space, i.e. to introduce a kind of fields
(cf. Sect.~\ref{external}).

Another possibility is to change the framework. For example, we
can stop here to deal with non-relativistic physics and recall
that it is energy rather than mass, which is conserved. However,
in this case even before discussing any mechanisms of the energy
transfer (which should be somewhat similar to the previous
consideration) we have to acknowledge, that we should immediately
change the basic equations of both kinematics (describing a motion
of objects) and dynamics (describing the gravity).

In other words, we cannot simply say that the masses are
time-dependent, we should go further to create a consistent
construction which allows such dependence within a certain
framework.

We have not discussed here two other problems. One is related to
what can be really measured. When we look for a change in the
distance, we usually mean that we look for a change in its
numerical value in some units. The interpretation would strongly
depend on what kind of clocks we use (the measurement of the
distance is usually a measurement of light-propagation time) and
on our assumptions on what can happen with the value of speed of
light $c$.

The other question is the gravitational constant, whether it can
change or we look for a variation of the masses only. To create a
`real' variation of $G$ we need to modify theory of
gravity. To make an `observable' variation of $G$, it is sufficient
either to change the units, or the masses, because we cannot
observe $G$ separately from gravitating masses and separately from
measuring masses and distances or related quantities in certain
units.

\subsection{Variations of the constants and violation of the
Lorentz invariance or LTI}

When searching for a variation of constants one has to remember
about a possible connection with various symmetries related to
relativity.

The simplest issue is an observational one. The variations are
long-term changes in values of fundamental constants, while a
violation of Lorentz invariance could produce periodic effects
because of the Earth rotation and its motion around the Sun (more
precisely both motions should be considered with respect to the
remote stars\footnote{Even that is not absolutely clear. There are
at least two preferred frames moving with respect to each other:
one is related to the local DM cluster, while the other is related
to the isotropic CMB. They suggest a different distance scale and
both can in principle lead to periodic variations. Any periodic
effect induced by the dark-matter-determined frame has no relation
to a violation of the Lorentz invariance. With the CMB that is not
clear. CMB proper is a kind of `environment'. Meanwhile, if there
is any fundamental violation of the Lorentz invariance, we would
expect that violation determined the frame where the Big Bang
happened and thus where the CMB is isotropic. So this frame is
specific because of a possible violation and because of
environmental effects, related to violations of the Lorentz
invariance in the remote past}). That can be resolved
experimentally.

The other issue is a reason for a variation of constants. There
are basically three options.
\begin{itemize}
\item Variation of constants could be caused by `long-range'
environment. An example is the phase transition during the early
time of the Universe. That has no relation to relativity.
\item There is a certain dynamics directly in space-time
continuum, which drives both: a violation of the relativity and a
variation of the constants. An example could be a consideration of
our 4-dimensional world as a result of compactification with the
radius of compactification dynamically changing.
\item An in-between option is a such kind of environment which
affects some relativity issues naively understood. For instance,
presence of a `medium' does not violate the relativity once we
speak about media as an non-fundamental issue added as an
environment. Meanwhile, we can choose to consider theory with media
as a fundamental `quasifree' theory with broken relativity. What
is important is the scale of phenomena. When we speak about the
propagation of light and the interaction of classical
macroscopic sources of the electric or magnetic field in a gas, we
deal with a kind of fundamental electromagnetic theory with
violated relativity. However, considering atomic spectra, we find
that they are related to electrodynamics of vacuum and all
deviations from the vacuum case happen on a certain macroscopic
distance scale.
\end{itemize}
Indeed, only the second option is related to a violation of local
position/time invariance.

\section{Planck scale physics in our low-energy world}

\subsection{Renormalization and Planck scale physics}

A big success of quantum electrodynamics was due to the
introduction of the renormalization scheme. Briefly speaking,
quantum electrodynamics (QED) is in a sense not a fundamental
theory, but a fundamental constraint.

A fundamental theory is such a theory that being formulated in
terms of certain laws and certain parameters produces a result in
terms of those fundamental parameters. Such a view on QED has
failed because of divergences.

Indeed, in reality everything in physics should be finite, but we
know that we possess only some knowledge on asymptotic low-energy
behavior of various physical quantities. Very often applying
asymptotics beyond their applicability one goes into unphysical
behavior of various results and, sometimes, to divergences. To
make divergences finite one has to use a complete description,
not its asymptotics, with exact laws instead of their asymptotic
forms.

The problem of QED is that we cannot learn anything about the
`complete description' and `exact laws', because they are related
to physics beyond our reach. Using different models for this
physics (i.e. different regularizations) we arrive at different
results.

Power of the renormalization procedure is in the treatment of QED
as a fundamental constraint, not as a theory. We can calculate a
long-range Coulomb-like interaction (which determines an
observable value of the electric charge), we can study electron's
kinetic (or complete) energy (which determines an observable value
of the electron's mass) and we can measure a number of other
properties such as the anomalous magnetic moment of an electron and
the Lamb shift in the hydrogen atom. The `constraint' means that
they are correlated and we can calculate the correlation. Learning
some of these values from experiment, we can predict the others.

The `fundamental constraint' means that it is enough to learn very
few values to predict all the others with an arbitrary accuracy
(or more precisely---as accurate as we can treat them as pure QED
values). For QED predictions, as we know, it is sufficient to
measure the elementary charge and masses of each kind of
particles.

The alternatives are known---to predict a value of the electric
or magnetic field of a non-elementary object we have to know not
only its charge and mass, but also all details of the distribution
of its electric and magnetic moments (and a number of parameters
beyond that). Those details should be also measured. So we need an
infinite number of the parameters.

Does the renormalization mean that the Planck scale does not
contribute to our experiments? No, it does not mean that. The
Planck scale indeed contributes, but it does not contribute to the
constraint, because it only affects values of masses and charges,
but we do not calculate, but measure them. That makes the
Planck-scale effects unobservable. To observe we should compare a
measurement and a calculation, but we have only results of
measurements. However, there is an option when we should be able
to see some effects of the Planck scale \cite{looking}. That is a
case when we have certain dynamic effects at the Planck scale
(e.g., a variation of some constants) or some violation of
symmetries which would make our low-energy picture wrong.

For instance, if we assume that we live in a multidimensional
world with a changing compactification radius, we may expect that
electron's mass and charge should vary. The effects depend on the
model of origin of bare masses and coupling constants. The bare
values can change or, alternatively, the bare values would stand
unchanged, while the renormalization term would change.

As an illustration we recall that we can see a number of
consequences of special relativity and quantum mechanics in
non-relativistic macroscopic phenomena. For instance, with
a precision achieved in the mass spectroscopy, we can see a
non-conservation of the mass because of the binding energy.
Various interferometers of the macroscopic scale prove that the
trajectory is not a well-defined property. And so on.

\subsection{The classical Michelson-Morley experiment and
calculability of the fine structure constant}

Here, we consider as an example a possible problem with an
interpretation of a classical version of the Michelson-Morley
experiment. In the experiment some pieces of bulk matter were
rotated. It was expected that when rotating their linear scale
would not change and comparing the light propagation in different
arms of the interferometer we can judge whether the speed of light
is the same in different directions.

Meanwhile, there is no just `speed of light' if we assume a
violation of the relativity. There are many different effects
instead. But still we can expect that non-relativistic physics
would not change too much. The size of a piece of atomic bulk
matter is basically determined by the non-relativistic Coulomb
interaction and we can believe that comparing a non-relativistic
distance and relativistic propagation of light we should have a
clear signature.

However, that is not that simple. The non-relativistic size
depends on the electron mass and its charge. Let us, e.g., assume
that $\alpha$ is calculable and that means that elementary charge
can be presented in terms of $\hbar c$. If special relativity is
violated and, e.g., $c=c(x)$, we should also arrive at $e(x)$.
(More precisely, we should speak about calculability of the
non-relativistic long-distance interaction of two charges {\em ab
initio\/}.) The Coulomb interaction, which is a pure
non-relativistic effect, would nevertheless be sensitive to a
violation of the special relativity. Rotating the interferometer
built as a bulk body we would deal with two effects: changes in
speed of propagation of light and in a distance between the
mirrors.

In other words, if the elementary charge is a fundamental quantity
which is not correlated with the speed of light, the Coulomb-law
energy is $E=Z_1Z_2e^2/r$ with possibly $e=e(t)$, while if
$\alpha$ is calculable, it is $E=Z_1Z_2\alpha\hbar c/r$ with
$\alpha=\alpha(t)$. (We remind that a real picture should be
somewhat more complicated---instead of varying constants we should
introduce some additional parameters and their derivatives (see
above)).

In a more complicated way similar reasons can be related to masses
of an electron and nucleon. The complicity is because we rather
expect that $\alpha$, if calculable, is calculable in a kind of
one-step action (with further renormalization), while for masses
we need to go step by step. For instance, for the electron we
should first understand the calculability of parameters of the
Higgs sector.

That means that for a proper interpretation of a
Michelson-Morley-like experiment with an interferometer built on an
atomic bulk matter we need to consider a dynamic model of
structure of this kind of matter with a possible violation of
relativity. The latter may involve the Planck scale effects, where
a certain relation on low-energy fundamental constants can be set.

\subsection{How to violate symmetries?}

There is a number of ways to violate some symmetries. The most
naive way it to violate such a symmetry directly. For example, the
masses of the up and down quarks violate a chiral symmetry of QCD.
That is the most natural way for classical physics. In the case of
quantum field theory, such a violation for the relativity and
related effects can most likely take form of an external field
(see Sect.~\ref{external} below). In particular, the spontaneous
breakdown of symmetries takes the form of certain external fields.

Quantum theory also opens  a number of other options (see, e.g.,
\cite{looking}). One of them is a so called anomaly. The violating
term is a purely quantum effect proportional to $\hbar$. It
appears because of singularities in original theory. While in the
classical case the theory is symmetrical under a number of
transformations, it is not possible to regularize all singular
operators to keep all classical symmetries. Some of them have to
be violated in the quantum-field case. The most well-known example
is a so called axial anomaly, which violates chiral symmetry even
for massless quarks.

A very remote analogy is conservation laws in classical and
quantum physics. Description of quantum mechanics in terms of
classical mechanics is not well defined, which happens because of
commutativity of classical values and non-commutativity of their
quantum analogs. We should regularize it and as a result part of
classical symmetries may be realized in such a way that some
conservation laws cannot be measured at all (e.g., conservation of
the angular momentum as a vector). That example turns our
attention to problem of observations.

Some effects may be a pure observational problem. We can
illustrate it by comparing conservations in classical and quantum
physics. We remark that we cannot check any conservation laws, but
only their consequences. From the point of view of classical
physics we expect that we can measure different components of
angular momentum and check at some time whether they have the same
values. From quantum physics we know that they would not have the
same value and that we can directly check only conservation of one
component of the angular momentum. Conservation of the angular
momentum as a vector can be checked via some consequences, but not
directly.

The problem is with commutations of different components of the
angular momentum. Meanwhile, it is expected that operators of
coordinates can be not-commutative in the quantum gravity. That
would produce certain observational effects for naive tests.

\section{External fields and related effects \label{external}}

\subsection{External field as a violation of relativity and CPT}

Even considering various violations, we basically expect that the
relativity, CPT and many other would-be violated invariances are
still present in a sense. Their violations used to be suggested in
the form of a kind of external field of a classical (caused by
matter or dark matter) or quantum (condensate) origin. We refer
here to such a field as a `violating field'. The violating field
can have a certain simple form in a specific frame and the result
in other frames can be found by an appropriate Lorentz
transformation.

It is very natural that most of such violating fields are very
similar to conventional fields such as scalar, electromagnetic,
gravitational etc. That is not a surprise, because if we like to
introduce both conventional and violating fields, we start from
designing a certain interacting term in the Lagrangian which obeys
all necessary symmetries. There are two basic differences between
`true' fields, which we used to deal with, and violating ones. The
former are somewhat universally coupled to many objects and they
are a result of certain sources existing in the case, or they are
quantized as photons. The violating fields have no sources, they
are background fields; and what is very important they are
somewhat selectively coupled to other objects. We know only one
kind of such a field, the Higgs field, which violates
SU(2)$\times$U(1) symmetry in the Stardard Model of the
electroweak interactions. It is also not-universally coupled to
the matter fields and as a result the masses of charged leptons
and quarks are all very different.

Let us also remind that a violation of CPT, most wanted by
experimentalists, is such a violation when mass and charge of
particle and antiparticle are not the same. To provide the
different charges would be a big problem, since it assumes a
non-conservation of the charge by producing a pair of particle and
antiparticle (the alternative is a photon with a very small but
not vanishing charge which is also not good). Possessing different
masses means, that while the mass of an electron is
\[
m_-=m-\delta m\;,
\]
the positron mass is \[m_+=m+\delta m\;.
\]
However, the same effects can be obtained if we assume that
\[
m_q=m+qeU\;,
\]
where $q$ is charge, equal to $\mp 1$, and the electric
potential $U$ is defined as $U=\delta m/e$.

Meanwhile, because of the gauge invariance we cannot observe any
constant homogenous potential since the related strength of the
electric field is zero. Does it mean that such a term is not
observable at all? The answer depends on how we treat different
particles and what kind of problem we study. If, e.g., we
consider the muon in the same way, but if two effective potentials
are not the same ($U_e\neq U_\mu$), we should be able to observe
their difference. The decay of muon and antimuon should have
slightly different kinematics and the difference in their
lifetime caused by the different phase volume of the decay
product, would be proportional to $U_e-U_\mu$. To understand that
we can have in mind so unrealistically large value of this
difference that a muon would decay, but an antimuon would not.

If we do parametrization more rigorously and introduce $\gamma_0$,
instead of $q$ (or more correctly to deal with the substitute
\[
m\to m+\gamma^\mu a_\mu\;,
\]
where $a_\mu$ is a time-like vector),
the result remains the same. An observable departure from CPT
should be proportional to a certain difference of parameters of
two particles, involved into calculations (cf. contributions of
the $a$ term in \cite{kostelecky}).

\subsection{`Selective' external fields and macroscopic experiment}

As we could see above, the violating term is similar to the electric
potential, but it is a kind of a selective field which should
interact differently with different kinds of particles and only
the differences can be observed.

A situation when the searched violating external fields are
similar to conventional electromagnetic fields, but to selective
ones, is very important from a practical point of view.

What is an electromagnetic field from a pragmatic point of view?
In conventional electrodynamics an electron, a proton, a muon etc.
sees the same electric field (once we neglect motional magnetic
effects). If we set a different background field for different
particles, that may well serve for producing a CPT violation or
violation of the Lorentz invariance. The conventional magnetic
field interacts universally with moving charges and there is an
additional interaction with spins or rather with related magnetic
moments. Some of spin magnetic moments are calculable {\em ab
initio\/} as for an electron or a muon, some should be treated
phenomenologically, as for a proton or a neutron.

Meanwhile, any experimental setup involves macroscopic bodies,
which can interact with the electromagnetic field, and some of
them do interact. Certain substances do that in peculiar ways,
when only one kind of universal interactions is involved.

For example, the solid conductors screen the `true' electric field
via a rearrangement of the electron density. With a violating
field, which interacts with the electrons, added, the conductors
should screen the field as seen by their electrons, i.e. they
screen both `true' electric field and the additional field
interacting with electrons. As a result they leave a certain
electric field inside the screened area. If the probe particle
will be an electron, no field would be seen, because the remaining
electric field will compensate the violating field. If the probe
particle is a proton, it should see a certain effective field
which is a difference of violating proton's and electron's fields.

A similar situation is with a magnetic-field-like violating field.
The `true' magnetic field interacts universally with all particles
and the same field is seen by any orbital and spin magnetic
moments in a consistent way. The violating field could be
different for different particles and it may interact in a
different way with the spin and the orbital motion. Some magnetic
screening materials act via a production of certain electron
currents (i.e., the orbital motion), while others via a
rearrangement of the electron spins. While providing the
screening, the electrons will act in such a way that they will
cancel all the field, including a violating component.

These examples show that while there may be certain vacuum
effects, the experiments are never done in vacuum. A
certain screening is always needed to avoid residual electromagnetic
fields. In the case of CPT violating fields, acting as `selective'
electromagnetic fields, a certain electron-interacting component
of such a field should be compensated by an electromagnetic field
created by the shielding material. That should be taken into
account for interpretation of such experiments.

\section{Microscopic and macroscopic description}

While a natural microscopic picture involves an effective external
field, the natural macroscopic description is rather a kind of
dilute medium (e.g., for the dark matter) which weakly interacts
with light etc. It is not the ether! The dilute medium
obviously affects the Doppler effect etc. and produces a signal for
the Michelson-Morley experiment. The speed of light would not be a
universal `c'. However, for microscopic properties such as a value
of $mc^2$ as the rest energy that would be different. Either they
would have no relation to measured velocity of light in the media,
or there would be different changes.

For example, considering the time dilation of the lifetime of an
unstable nuclear level we should consider the nucleus which
lifetime changes because of two different effects: conventional
Lorentz transformation and interaction with the dilute media
particle. The same should be with various ratios of different
transition frequencies from the same atom.

Indeed, the particle interaction with the dilute media depends on
their relative velocity but also on various other parameters.
E.g., we can assume that the particle directly interacts with the
(dark-matter) medium (via a heavy intermediate boson) and the
interaction with light is indirect and somewhat weaker. Or on the
contrary we can suggest that the dilute medium is weakly coupled
to the light directly and any interaction to the other matter is
an induced effect. Indeed, with a fixed value of the light-media
interaction, the effects of matter-media interaction can easily
vary by orders of magnitude.

The crucial point here is a possible scale of the effects. The
Michelson-Morley experiment and some others are of macroscopic
nature and they can check various symmetries on a large scale with
respect to atomic and particle effects scale. The latter scale
could be studied via a different kind of experiments and it is not
necessary that the result be consistent.

Addressing different scales of times and distances we study a
part of effects and trying to generalize the results may do some
model-dependent suggestions.

\section{Summary}

Above we have demonstrated that looking for some new physics and
in particular for possible violations of some symmetries it is
hard to avoid certain model dependence which may sometimes produce
a misleading interpretation. It is hard to give any general
advices except for being careful.

This work was supported in part by the RFBR grant \# 06-02-04018
and DFG grant GZ 436 RUS 113/769/0.

\end{document}